# Weakness in a Mutual Authentication Scheme for Session Initiation Protocol using Elliptic Curve Cryptography

Debiao He

School of Mathematics and Statistics, Wuhan University,

Wuhan, People's Republic of China

hedebiao@163.com

*Abstract*—The session initiation protocol (SIP) is a powerful signaling protocol that controls communication on the Internet, establishing, maintaining, and terminating the sessions. The services that are enabled by SIP are equally applicable in the world of mobile and ubiquitous computing. In 2009, Tsai proposed an authenticated key agreement scheme as an enhancement to SIP. Very recently, Arshad et al. demonstrated that Tsai's scheme was vulnerable to offline password guessing attack and stolen-verifier attack. They also pointed that Tsai's scheme did not provide known-key secrecy and perfect forward secrecy. In order to overcome the weaknesses, Arshad et al. also proposed an improved mutual authentication scheme based on elliptic curve discrete logarithm problem for SIP and claimed that their scheme can withstand various attacks. In this paper, we do a cryptanalysis of Arshad et al.'s scheme and show that Arshad et al.'s scheme is vulnerable to the password guessing attack.

*Keywords- Authentication; Elliptic curve cryptosystem; Security; Session initiation protocol*

## I. INTRODUCTION

In today's and future wired or wireless networks, multimedia service is a great importance application class. Especially, the next generation of wireless networks will be based on all-IP architecture. One of the most important protocols supporting multimedia services is the session initiation protocol (SIP) [1] . In 1999, the Internet Engineering Task Force (IETF) proposed the SIP for the IP-based telephony protocol [2-7]. Because SIP is a text-based peer-to-peer protocol, it uses Internet protocols such as hyper text transport protocol (HTTP) and simple mail transport protocol (SMTP) [8] . When a user wants to access an SIP service, he/she has to perform an authentication process in order to get various services from the remote server. Therefore, the security of SIP is becoming more important [4],[6],[8] and the SIP authentication scheme is the most important issue for SIP. However, the original authentication scheme for SIP does not provide strong security because it is based on the HTTP digest authentication [2]. The services that are enabled by SIP are equally applicable in the world of mobile and ubiquitous computing. For example, a user registers his/her location with a SIP server and then the server knows if the user is available and where the user can be found. In addition, the location could be home, work, or mobile. Now, SIP has been defined by the 3GPP (the 3rd Generation Partnership Project) as a signal protocol for the third-generation communication system [9].

Yang et al. [10] pointed out that the procedure of the original SIP authentication scheme based on HTTP digest authentication is vulnerable to the off-line password guessing attack and the server spoofing attack. They also proposed a secure authentication scheme for SIP to resist the attacks. However, the computational cost of Yang et al.'s scheme is very high, making it unsuitable for practical applications. To improve the performance, Durlanik et al. [11] proposed an authentication scheme using the elliptic curve cryptography(ECC)[12, 13]. Later, Tsai [14] proposed an efficient nonce-based authentication scheme. Since all the communication messages are encrypted/decrypted by using one-way hash function and XOR operation, its computation cost is low, making it promising for low-power processors. Unfortunately, Arshad et al.[15] found Tsai's scheme[14] was still vulnerable to offline password guessing attack and stolen-verifier attack while. They also demonstrated that Tsai's scheme did not provide known-key secrecy and perfect forward secrecy[9]. To overcome the weaknesses, Arshad et al.[15] also proposed an improved authentication using ECC for SIP.

In this letter, we show that Arshad et al.'s scheme may suffer from off-line password guessing attack if only one elliptic curve scalar multiplication is required in the generation of password verifier from password. The rest of the paper is organized as follows. Section 2 gives the review of the Arshad et al.'s scheme. Section 3 discusses the cryptanalysis of Arshad et al.'s scheme. Finally, we conclude the paper in Section 4.

II. REVIEW OF ARSHAD ET AL.'S SCHEME FOR SIP

In this section, we briefly review Arshad et al.'s nonce-based authentication scheme for SIP [1]. There are two phases in Arshad et al.'s scheme: registration and authentication. Notations used in this paper are defined as follows.

- $U$ : the client user;
- $S$ : the server;
- $D$ : a uniformly distributed dictionary;
- $PW$ : a low-entropy password of U extracted from $D$;

- $K_S$: a high-entropy secret key of $S$, which is only known by the server and must be safeguarded;
- $SK$: a shared common session key between $U$ and $S$;
- $p, n$: two large prime numbers;
- $F_p$: a finite field;
- $E$: an elliptic curve defined on finite field $F_p$ with large order;
- $G$: the group of elliptic curve points on $E$;
- $P$: a point on elliptic curve $E$ with order $n$;
- $h(\cdot)$: a secure one-way hash function;
- $\|$: a string concatenation operation
- $\oplus$: a string XOR operation

A. *Registration phase*

When $U$ wants to register and become a new legal user, $U$ and $S$ execute the following steps over a secure channel.

1) $U \to S : (username, PW)$

   $U$ submit his $username$ and $PW$ to $S$. $S$ computes two secret values $HPW = h(username \| PW)$ and $HK_S = h(username \| K_S)$.

2) $S$ computes the password verifier $VPW = HPW \oplus HK_S$ for $U$.

3) $S$ stores $U$'s $username$ and $VPW$ in the user account database.

B. *Authentication phase*

If a legal user wants to login into $S$, he must type his $username$ and $PW$. All steps of authentication phase, as shown in Fig. 1, executed as follows.

1) $U \to S :$ REQUEST ($username$, $R_1$)

   $U$ computes $HPW = h(username \| PW)$. Then $U$ generates a random number $r_1 \in Z_n^*$, computes $R_1 = (HPW \cdot r_1) \times P$ and sends a request message as REQUEST ($username$, $R_1$) to $S$.

2) $S \to U :$ CHALLENGE ($realm$, $R_2$, $h_1$)

Upon receiving the request message, $S$ extracts $HPW$ from $VPW$ by computing $HK_S = h(username \| K_S)$ and $HPW = VPW \oplus HK_S$, where $VPW$ is a stored password verifier for $U$ in the user account database. Then, $S$ computes $R_1' = HPW^{-1} \times R_1$. Now, $S$ generates a random number $r_2 \in Z_n^*$, and computes $R_2 = r_2 \times P$, $SK_S = r_2 \times R_1'$ and $h_1 = h(SK_S \| R_2)$. At last, $S$ sends a challenge message CHALLENGE ($realm, R_2, h_1$) to $U$.

3) $U \to S$ : RESPONSE ($username, realm, h_2$)

Upon receiving the challenge message, $U$ computes $SK_U = r_1 \times R_2$ and checks whether the equation $h_1 = h(SK_U \| R_2)$ holds. If the equation does not hold, $U$ rejects the server challenge message. Otherwise, $U$ authenticates $S$ and computes $h_2 = h(username \| realm \| SK_U)$. Finally, $U$ sends a response message RESPONSE ($username, realm, h_2$).

4) Upon receiving the response message, $S$ computes $h(username \| realm \| SK_S)$ and verifies whether it is equal to the received response $h_2$. If they are not equal, $S$ rejects the user response message. Otherwise, $S$ authenticates $U$ and accepts the user's login request.

After mutual authentication between $U$ and $S$, $SK = SK_U = SK_S = r_1 r_2 \times P$ is used as a shared session key.

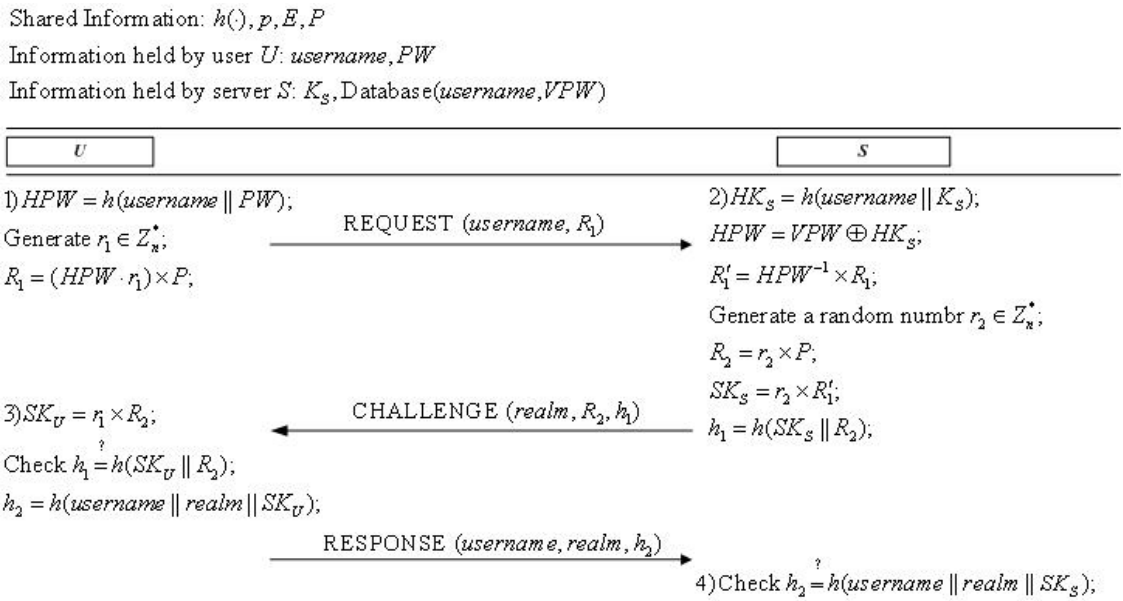

Fig. 1. Authentication phase of Arshad et al.'s scheme

## III. CRYPTANALYSIS OF ARSHAD ET AL.'S SCHEME FOR SIP

Off-line password guessing attack succeeds when there is information in communications, which can be used to verify the correctness of the guessed passwords. In [15], Arshadet et al. claimed that their protocol can resist the off-line password guessing attack. However, in this section, we will show that the off-line password guessing attack, not as they claimed, is still effective in Arshadet et al.'s protocol. Our attack consists of two phases.

**Phase 1**:

1) The adversary $A$ chooses a random number $r_1 \in Z_n^*$, computes $R_1 = r_1 \times P$, then impersonates $U$ and sends a request message REQUEST ($username$, $R_1$) to $S$.

2) Upon receiving the request message, $S$ extracts $HPW$ from $VPW$ by computing $HK_S = h(username \| K_S)$ and $HPW = VPW \oplus HK_S$, where $VPW$ is a stored password verifier for $U$ in the user account database. Then, $S$ computes $R_1' = HPW^{-1} \times R_1 = HPW^{-1} \cdot r_1 \times P$. Now, $S$ generates a random number $r_2 \in Z_n^*$, and computes $R_2 = r_2 \times P$, $SK_S = r_2 \times R_1' = HPW^{-1} \cdot r_1 \cdot r_2 \times P$ and $h_1 = h(SK_S \| R_2)$. At last, $S$ sends a challenge message CHALLENGE ($realm$, $R_2$, $h_1$) to the adversary $A$.

**Phase 2**:

The adversary $A$ then launches the off-line password guessing attack as follows:

1) $A$ guesses a candidate passwords $\overline{PW}$, and computes $\overline{HPW} = h(username \| \overline{PW})$.

2) $A$ computes $\overline{SK_S} = \overline{HPW}^{-1} \cdot r_1 \times R_2$.

3) $A$ checks whether $h_1$ and $h(\overline{SK_S} \| R_2)$ are equal. If $h_1$ and $h(\overline{SK_S} \| R_2)$ are equal, the adversary can conclude that $U$'s password $PW = \overline{PW}$. Otherwise, adversary repeat steps 1), 2) and 3) until the correct password is found.

From the above description, we know the adversary can get the password. Therefore, Arshad et al.'s scheme is vulnerable to the off-line password guessing attack.

## IV. CONCLUSIONS

In this letter, we have shown that Arshad et al.'s authentication scheme for session initiation protocol is vulnerable to password guessing attack. The analysis shows Arshad et al.'s authentication scheme is not for practical application.

# REFERENCES

[1] D. Geneiatakis, T. Dagiuklas, G. Kambourakis, C. Lambrinoudakis, S. Gritzalis, and S. Ehlert. "Survey of security vulnerabilities in session initiation protocol," IEEE Communications Surveys and Tutorials, vol. 8, no. 3, pp. 68-81, Jul. 2006.

[2] J. Franks, P. Hallam-Baker, J. Hostetler, S. Lawrence, P. Leach, and A. Luotonen. HTTP Authentication: Basic and Digest Access Authentication. IETF RFC2617, Jun. 1999.

[3] M. Handley, H. Schulzrinne, E. Schooler, and J. Rosenberg. SIP: Session Initiation Protocol. IETF RFC2543, Mar. 1999.

[4] M. Thomas. SIP Security Requirements. IETF Internet Draft (draftthomas-sipsec-reg-00.txt), Work In Progress, Nov. 2001.

[5] J. Rosenberg, H. Schulzrinne, G. Camarillo, A. Johnston, J. Peterson, and R. Sparks. SIP: Session Initiation Protocol. IETF RFC3261, Jun. 2002.

[6] J. Arkko, V. Torvinen, G. Camarillo, A. Niemi, and T. Haukka. Security Mechanism Agreement for SIP Sessions. IETF Internet Draft (draft-ietf-sipsecagree-04.txt), Jun. 2002.

[7] A. B. Johnston. SIP: Understanding the Session Initiation Protocol. 2nd ed. Artech House; 2004.

[8] L. Veltri, S. Salsano, and D. Papalilo. "SIP security issues: the SIP authentication procedure and its processing load," IEEE Network, vol. 16, no. 6, pp. 38-44, 2002.

[9] M. G. Martin, E. Henrikson, and D. Mills. Private Header (P-Header) Extensions to the Session Initiation Protocol (SIP) for the 3rd-Generation Partnership Project(3GPP). IETF RFC3455. 2003.

[10] Yang CC, Wang RC, Liu WT (2005) Secure authentication scheme for session initiation protocol. Comput Secur 24:381–386

[11] Durlanik A, Sogukpinar I (2005) SIP Authentication Scheme using ECDH. World Enformatika Society Trans Eng Comput Technol 8:350–353

[12] Miller V.S. (1985) Use of elliptic curves in cryptography. In: Proceedings of the Advances in Cryptology-CRYPTO'85, New York, USA, 417–426.

[13] Koblitz N. (1987) Elliptic curve cryptosystem, Mathematics of Computation 48: 203–209.

[14] Tsai JL (2009) Efficient nonce-based authentication scheme for session initiation protocol. Int J Netw Secur 8(3):312–316

[15] Arshad R., Ikram N., Elliptic curve cryptography based mutual authentication scheme for session initiation protocol, Multimedia Tools and Applications, DOI: 10.1007/s11042-011-0787-0